\documentclass[runningheads]{llncs}
\usepackage[T1]{fontenc}
\usepackage{amsmath}
\usepackage{amsfonts}
\usepackage{dsfont}
\usepackage{graphicx}
\usepackage{tikz}
\usepackage{multirow}
\usepackage{array}
\usepackage{tabularx,booktabs}
\usepackage{subcaption}
\usepackage{dirtytalk}
\usepackage{bm}
\usepackage{dsfont}
\usepackage{float}
\usetikzlibrary{arrows.meta}
\usepackage{color}
\usepackage{hyperref}
\usepackage{rotating}
\usepackage[misc]{ifsym}

\begin{document}

\title{SpurBreast: A Curated Dataset for Investigating Spurious Correlations in Real-world\\Breast MRI Classification}

\titlerunning{SpurBreast: A Curated Dataset for Investigating Spurious Correlations}

\author{
Jong Bum Won\inst{1} \and
Wesley De Neve\inst{1,2} \and
Joris Vankerschaver\inst{1,3} \and\\
Utku Ozbulak\inst{1,2} (\Letter)
}

\authorrunning{Won et al.}

\institute{
Center for Biosystems and Biotech Data Science, Ghent University Global Campus, Incheon, Republic of Korea \and
IDLab, ELIS, Ghent University, Ghent, Belgium \and
Department of Applied Mathematics, Computer Science and Statistics, Ghent University, Ghent, Belgium\\
 (\Letter) \email{utku.ozbulak@ghent.ac.kr}
}

\maketitle
\begin{abstract}
\let\thefootnote\relax\footnotetext{Accepted for publication in the 28th International Conference on Medical Image Computing and Computer Assisted Intervention (MICCAI), 2025.}
Deep neural networks (DNNs) have demonstrated remarkable success in medical imaging, yet their real-world deployment remains challenging due to spurious correlations, where models can learn non-clinical features instead of meaningful medical patterns. Existing medical imaging datasets are not designed to systematically study this issue, largely due to restrictive licensing and limited supplementary patient data. To address this gap, we introduce SpurBreast, a curated breast MRI dataset that intentionally incorporates spurious correlations to evaluate their impact on model performance. Analyzing over 100 features involving patient, device, and imaging protocol, we identify two dominant spurious signals: magnetic field strength (a global feature influencing the entire image) and image orientation (a local feature affecting spatial alignment). Through controlled dataset splits, we demonstrate that DNNs can exploit these non-clinical signals, achieving high validation accuracy while failing to generalize to unbiased test data. Alongside these two datasets containing spurious correlations, we also provide benchmark datasets without spurious correlations, allowing researchers to systematically investigate clinically relevant and irrelevant features, uncertainty estimation, adversarial robustness, and generalization strategies. Models and datasets are available at \href{https://github.com/utkuozbulak/spurbreast}{github.com/utkuozbulak/spurbreast}.
\end{abstract}


\section{Introduction}

Deep Neural Networks (DNNs) have achieved remarkable success in medical imaging, demonstrating the potential to match or even surpass expert performance in diagnosing diseases \cite{pham2021derma,wu2020breast}. Despite these advancements, their deployment in clinical settings remains challenging due to their susceptibility to distribution shifts -- where differences between training and real-world data lead to significant drops in performance \cite{saab2022reduce,wiens2019}. A critical factor contributing to this issue is the presence of spurious correlations, which occur when models inadvertently learn associations between irrelevant features and target labels, instead of focusing on clinically meaningful patterns \cite{hermann2024on}. In medical imaging, such correlations can arise from demographic biases, scanner artifacts, or variations in clinical settings, leading to models that fail to generalize effectively and, in turn, pose substantial risks in real-world applications \cite{kocak2024}. These unintended dependencies not only reduce diagnostic accuracy but also have broader implications, such as influencing clinical decision-making processes and potentially introducing biases in healthcare access and insurance claims.

Although widely-used medical imaging datasets such as CheXpert, MURA, and MIMIC-CXR have facilitated the development of AI models \cite{chexpert,mura,mimic_cxr}, they are not specifically designed to investigate spurious correlations. Existing datasets tailored for this purpose, such as ImageNet-C/P and Spawrious \cite{hendrycks2018benchmarking,anonymous2024spawrious}, primarily focus on natural and synthetic images, which fail to capture the unique complexities of medical imaging.

Creating curated datasets in medical imaging containing well-documented spurious correlations presents unique problems, primarily due to licensing restrictions and regulatory guidelines that necessitate the use of de-identified and unaltered medical images \cite{clark2013,johnson2016}. While synthetic datasets have been proposed to circumvent these issues, they often lack realism and fail to capture the variability inherent in real-world medical imaging~\cite{stanley2025and}. On the other hand, discovering naturally occurring spurious correlations is particularly difficult because they often stem from subtle and indirect relationships~\cite{olesen2024slicing}. Furthermore, the presence of domain-specific biases and imbalanced data distributions exacerbates the problem, as certain demographic groups or disease types may be overrepresented, making it challenging to disentangle spurious associations from meaningful clinical features~\cite{larrazabal2020gender,vaidya2024demographic}.

In this work, we introduce \textbf{SpurBreast}, a curated dataset designed to study spurious correlations in real-world breast MRI data. It consists of real-world patient data, carefully curated to include well-documented spurious correlations such as those related to patient demographics and imaging equipment. In addition, we have conducted an extensive experimental analysis on supplementary patient data to systematically assess the impact of spurious correlations on model performance. Unlike existing datasets, SpurBreast provides a comprehensive framework that allows researchers to study the influence of spurious correlations under controlled conditions, facilitating the development of more robust and generalizable AI models for medical imaging.

\begin{figure}[t!]
\centering
\begin{subfigure}{0.25\textwidth}
\centering
\includegraphics[width=0.95\textwidth]{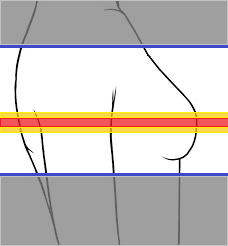}
\caption{Side profile}
\label{fig:MRI_chest_drawing}
\end{subfigure}
\centering
\begin{subfigure}{0.74\textwidth}
\includegraphics[width=0.17\textwidth]{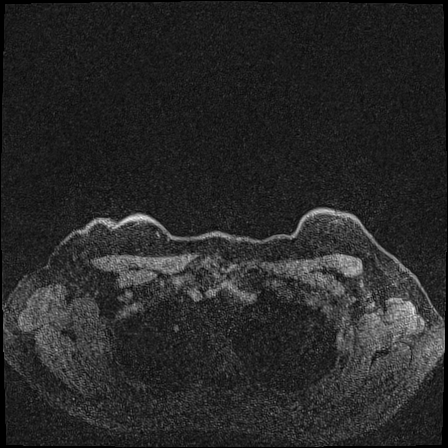}
\includegraphics[width=0.17\textwidth]{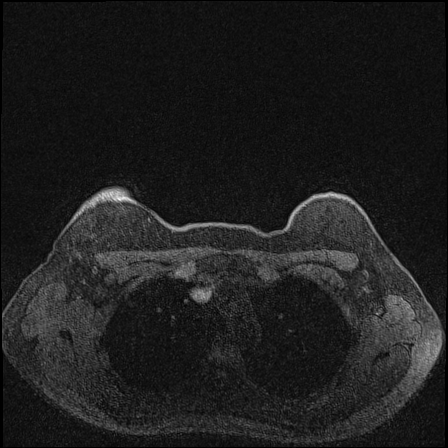}
\includegraphics[width=0.17\textwidth]{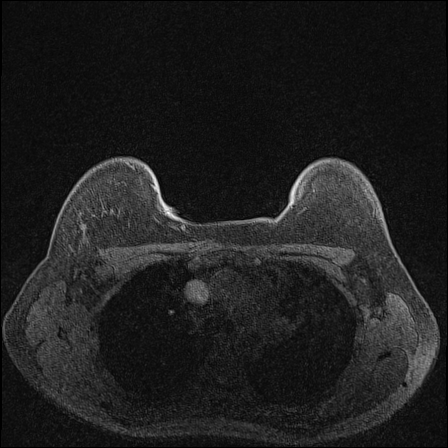}
\includegraphics[width=0.17\textwidth]{pat_103/1-029-103.png}
\includegraphics[width=0.17\textwidth]{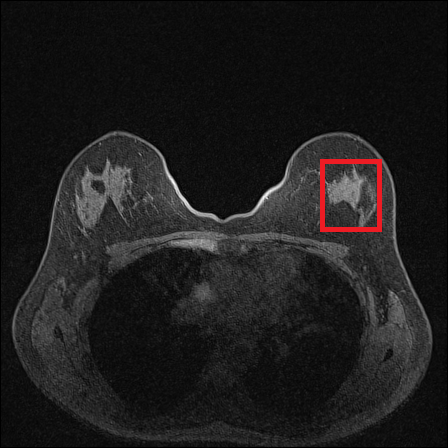}
\\
\includegraphics[width=0.17\textwidth]{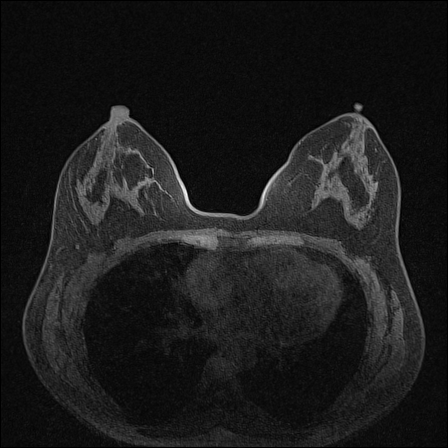}
\includegraphics[width=0.17\textwidth]{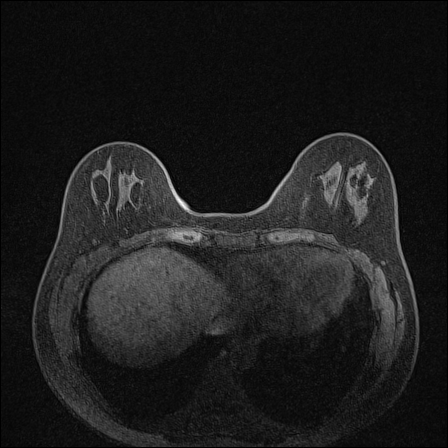}
\includegraphics[width=0.17\textwidth]{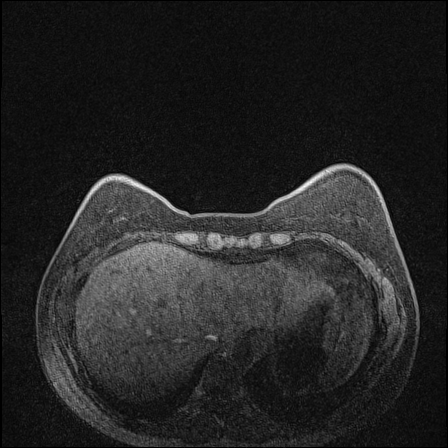}
\includegraphics[width=0.17\textwidth]{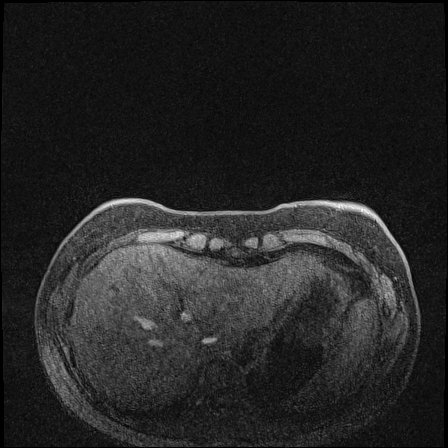}
\includegraphics[width=0.17\textwidth]{pat_103/1-133-103.png}
\caption{Example MRI slices}
\label{fig:MRI_slices}
\end{subfigure}
\caption{(a) A side profile diagram of the breast, highlighting the imaging region. Slices in the red area contain MRI images with breast tumors, slices in the yellow area are buffer zones and are not used, and slices in the white region do not contain invasive breast tumors. (b) Example MRI slices obtained from the specified cross-sectional region. Image with the highlighted red box in one slice indicates an invasive breast tumor.}
\label{fig:data_overview}
\end{figure}

\section{Data}
\label{sec:data}

\textbf{MRI images}.\,The data used in this study are from the DUKE Breast Cancer Dataset~\cite{duke_dataset}, a comprehensive single-institutional retrospective collection of 3D MRI scans from over 900 patients with biopsy-confirmed invasive breast cancer at a university hospital. Each study includes a 3D MRI acquired using 1.5T or 3T scanners, from patients in the prone position. On average, each 3D scan consist of 250 2D slices (see Figure~\ref{fig:data_overview}). For the predictive tasks, the slices are categorized into two groups: those containing breast tumors and those without. Following the approach of \cite{konz2022,kang2024exploring}, we establish a buffer zone between slices containing tumors and those that do not (highlighted in yellow in Figure~\ref{fig:MRI_chest_drawing}). Images within this buffer zone are excluded from analysis, and the remaining slices are labeled and used for the predictive task.

\textbf{Supplementary information}.\,Alongside the image data, separate tabular data cover various types of patient information including demographic, clinical, pathology, treatment information gathered from clinical notes, radiology reports, and pathology reports. Apart from that, these tabular data also encompass details about imaging devices and characteristics, including size, shape, texture, and enhancement patterns of both the tumor and surrounding tissue. Overall, these data contain more than 100 features. Unfortunately, the majority of these features are imbalanced in distribution, making the process of dataset creation challenging when taking those features into account.

\section{Methodology}

\subsection{Discovering Spurious Correlations}

\begin{figure}[t!]
\centering
\begin{subfigure}{0.45\textwidth}
\includegraphics[width=0.99\textwidth]{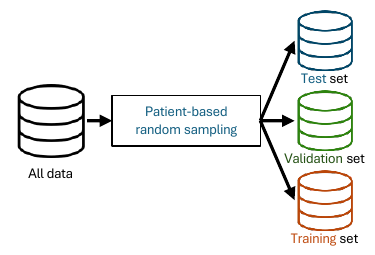}
\caption{Random split}
\label{fig:typical_split}
\end{subfigure}
\begin{subfigure}{0.52\textwidth}
\includegraphics[width=0.99\textwidth]{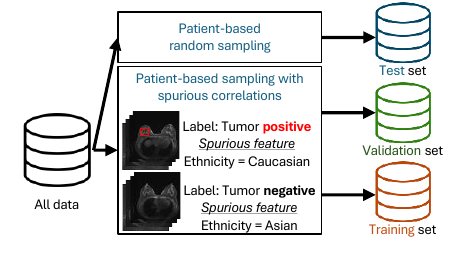}
\caption{Split with spurious correlations}
\label{fig:spurious_split}
\end{subfigure}
\caption{Illustration of the dataset creation process for discovering spurious correlations. (a) A typical patient-based random sampling approach, where the dataset is split into training, validation, and test sets to prevent overlap and ensure unbiased evaluations. (b) A modified sampling strategy where specific spurious correlations between predictive labels (tumor-positive and tumor-negative) and supplementary features (e.g., ethnicity) are deliberately introduced to study their effects on model performance.}
\label{fig:spurious_or_not}
\end{figure}

In typical machine learning studies, the data are split into training, testing, and validation subsets randomly in order to prevent spurious correlations from arising (see Figure~\ref{fig:typical_split}). Then, the training set is used to optimize the model parameters, the validation set to tune hyperparameters and guide model selection, and the testing set to serve as a final evaluation metric for generalization on new data.

Different than the aforementioned approach, in this study, we aim to find features that spuriously correlate with predictive labels and create a well-documented dataset containing clearly defined spurious correlations. To achieve this, we adopt the approach detailed below.

\textbf{Testing dataset}. Before creating any training and validation datasets, we randomly select 150 patients and use the tumor-positive and tumor-negative slices from those images for the testing dataset. Throughout the manuscript, we use this dataset to consistently measure the generalization performance of trained models on unseen data and ensure that measurements on the test data are easily comparable.

\textbf{Training and validation datasets}. Using the supplementary features described in Section~\ref{sec:data}, we divide the training and validation data, at the patient level, in a such way that the positive labels in both datasets are associated exclusively with images possessing a specific property, while the negative labels are linked to images with a different property. For example, using the ethnicity feature, we select all tumor-positive images from Caucasian patients, whereas all tumor-negative images are selected from Asian patients (see \figurename~\ref{fig:spurious_split}). This setup introduces a spurious correlation between ethnicity and the predictive label. If the spurious correlation is strong enough, models may exploit it to achieve high training and validation performance. However, their performance on the testing dataset, evaluated on data without spurious correlations, will be significantly lower.

\subsection{Datasets and Evaluated Features}

Using the approach outlined above, we create datasets based on a variety of unique features in the supplementary data. However, a substantial challenge in this process is data imbalance. For instance, only a small subset of patients in our dataset exhibit nipple retraction. Consequently, based on this feature, patients cannot be effectively split into training and validation sets to train DNNs, as the sample size is insufficient for meaningful training. Another challenge is missing data; a large number of features have a high proportion of missing labels, rendering them impractical for use in dataset splits. Based on the available data, we investigate more than 100 features but only report on several interesting and relevant features due to space constraints:

\textbf{Ethnicity}. This feature describes the self-reported ethnic background of patients, such as Caucasian, Asian, or African American. Differences in representation across groups can introduce spurious correlations if ethnicity disproportionately aligns with specific labels.

\textbf{Menopause status}.\,This feature indicates a patient’s menopausal status (e.g., premenopausal or postmenopausal), which affects breast tissue density. Its link to age and demographics may unintentionally correlate with diagnostic labels.

\textbf{Magnetic field strength}.\,The strength of the MRI scanner's magnetic field, measured in Tesla (e.g., 1.5T or 3T). Differences in field strength can affect image quality and introduce unintended correlations with labels.

\textbf{Surgery type}.\,This feature describes the type of surgery that will be performed on the patient (e.g., mastectomy or lumpectomy). Variability in surgical decisions may correlate with disease characteristics, leading to potential biases.

\textbf{Vertical alignment}.\,Indicates whether an image was vertically flipped during data augmentation. Uneven application of this transformation can create unintended associations with specific labels.

\textbf{Baseline performance}.\,Apart from the dataset splits created based on the aforementioned features, we also use three baseline performance datasets to evaluate the models on randomly sampled datasets without spurious correlations. These datasets are sampled based on patient counts and represent low-data, medium-data, and large-data benchmarks.

Details about these datasets, the number of images in the training, validation, and testing datasets, as well as the number of patients containing information regarding these features, are provided in Table~\ref{tbl:dataset_details}.

\begin{table}[t]
\centering
\scriptsize
\caption{The table presents the number of patients and images allocated to training, validation, and testing sets for different dataset configurations. The baseline datasets (low, medium, and large) contain randomly sampled patient data without spurious correlations, while the experimental datasets introduce specific spurious correlations based on features such as ethnicity, menopause status, MRI field strength, surgery type, and image alignments. The patient counts and corresponding image counts are shown for each dataset category.}
\label{tbl:dataset_details}
\begin{tabular}{lcccccc}
\toprule
\multirow{2}{*}{Feature} & \multicolumn{3}{c}{Patients in dataset} & \multicolumn{3}{c}{Images in dataset} \\
\cmidrule[0.5pt]{2-7}
~ & Training & Validation & Testing & Training & Validation & Testing \\
\midrule
Baseline (Low-data)  & 150 & 150 & \multirow{9}{*}{150} & 6,490 & 8,032 & \multirow{9}{*}{6,788}\\
Baseline (Medium-data) & 400 & 150 & ~ & 19,252 & 8,032 & \\
Baseline (Large-data) & 600 & 150 & ~ & 29,196 & 8,032 & \\
\cmidrule[0.5pt]{1-1}
Ethnicity & 200 & 100 & ~ & 4,974 & 2,408 & \\
Menopause & 400 & 150 & ~ &  10,488 & 4,330 & \\
Magnetic field strength & 400 & 150 & ~ & 9,562 & 3,576 &  \\
Surgery type & 400 & 150 & ~ & 7,860 & 2,916 & \\
Vertical alignment & 500 & 150 & ~ & 24,138 & 6,926 \\
\bottomrule
\end{tabular}
\end{table}

\subsection{Models}
\label{sec:models}

In the upcoming experiments, we use two models: a robust and widely adopted convolutional model in the literature (ResNet-50) \cite{resnet50}, and a transformer-based architecture, Vision Transformer (ViT-B/16)~\cite{vitb16}, which has demonstrated state-of-the-art performance in various computer vision tasks. All models are initialized with pretrained weights from the ImageNet dataset \cite{imagenet}, ensuring a strong starting point for transfer learning. The pretrained weights for these models are taken from the PyTorch library.

\begin{table}[t]
\centering
\scriptsize
\caption{Accuracy (Acc.), positive predictive value (PPV), and negative predictive value (NPV) for ResNet-50 and Vit-B models across training, validation, and testing datasets are provided, considering various experimental conditions, including randomized baseline data, demographic factors, and data augmentation techniques.}
\label{tbl:results}
\begin{tabular}{cc|ccc|ccc|ccc}
\toprule
~ & ~ & \multicolumn{3}{c}{Training} & \multicolumn{3}{c}{Validation} & \multicolumn{3}{c}{Testing} \\
\cmidrule[1pt]{3-11}
Feature & \phantom{-----}Model\phantom{-----} & \phantom{-}Acc.\phantom{-} & \phantom{-}PPV\phantom{-} & \phantom{-}NPV\phantom{-} & \phantom{-}Acc.\phantom{-} & \phantom{-}PPV\phantom{-} & \phantom{-}NPV\phantom{-} & \phantom{-}Acc.\phantom{-} & \phantom{-}PPV\phantom{-} & \phantom{-}NPV\phantom{-} \\
\midrule
\multirow{2}{*}{\shortstack{Baseline\\(Low-data)}} & ResNet-50 & 0.77 & 0.69 & 0.84 & 0.71 & 0.76 & 0.66 & 0.77 & 0.78 & 0.75 \\
~ & Vit-B & 0.79 & 0.74 & 0.85 & 0.75 & 0.70 & 0.80 & 0.79 & 0.70 & 0.88 \\
\cmidrule[0.5pt]{1-11}
\multirow{2}{*}{\shortstack{Baseline\\(Medium-data)}} & ResNet-50 & 0.76 & 0.66 & 0.86 & 0.75 & 0.77 & 0.73 & 0.77 & 0.77 & 0.77 \\
~ & Vit-B & 0.79 & 0.72 & 0.86 & 0.76 & 0.67 & 0.86 & 0.81 & 0.69 & 0.94 \\
\cmidrule[0.5pt]{1-11}
\multirow{2}{*}{\shortstack{Baseline\\(Large-data)}} & ResNet-50 & 0.80 & 0.74 & 0.86 & 0.76 & 0.76 & 0.77 & 0.82 & 0.78 & 0.86 \\
~ & Vit-B & 0.81 & 0.75 & 0.87 & 0.79 & 0.71 & 0.87 & 0.83 & 0.72 & 0.94 \\
\cmidrule[1pt]{1-11}
\multirow{2}{*}{Ethnicity} & ResNet-50 & 0.97 & 0.96 & 0.97 & 0.85 & 0.89 & 0.81 & 0.72 & 0.75 & 0.68 \\
~ & Vit-B & 0.83 & 0.81 & 0.86 & 0.81 & 0.77 & 0.85 & 0.71 & 0.65 & 0.77 \\
\cmidrule[0.5pt]{1-11}
\multirow{2}{*}{\shortstack{\textbf{{Magnetic}}\\\textbf{{Field Strength}}}} & ResNet-50 & \textbf{0.99} & 0.98 & 0.99 & \textbf{0.99} & 1.00 & 0.98 & \textbf{0.52} & 0.62 & 0.41 \\
~ & Vit-B & \textbf{0.98} & 0.98 & 0.98 & \textbf{0.99} & 0.99 & 0.99 & \textbf{0.55} & 0.66 & 0.43 \\
\cmidrule[0.5pt]{1-11}
\multirow{2}{*}{Menopause} & ResNet-50 & 0.91 & 0.91 & 0.92 & 0.85 & 0.88 & 0.82 & 0.71 & 0.77 & 0.65 \\
~ & Vit-B & 0.67 & 0.63 & 0.72 & 0.71 & 0.62 & 0.80 & 0.69 & 0.62 & 0.75 \\
\cmidrule[0.5pt]{1-11}
\multirow{2}{*}{Surgery Type} & ResNet-50 & 0.53 & 0.49 & 0.57 & 0.70 & 0.79 & 0.61 & 0.74 & 0.80 & 0.68 \\
~ & Vit-B & 0.80 & 0.73 & 0.87 & 0.75 & 0.63 & 0.87 & 0.77 & 0.66 & 0.88 \\
\cmidrule[0.5pt]{1-11}
\multirow{2}{*}{\shortstack{{\textbf{Vertical}}\\\textbf{{Alignment}}}} & ResNet-50 & \textbf{0.99} & 0.99 & 0.99 & \textbf{1.00} & 1.00 & 1.00 & \textbf{0.52} & 0.04 & 1.00 \\
~ & Vit-B & \textbf{0.98} & 0.98 & 0.98 & \textbf{1.00} & 1.00 & 1.00 & \textbf{0.52} & 0.05 & 1.00 \\
\bottomrule
\end{tabular}
\end{table}

\section{Experimental Results}
\label{sec:experimental_results}

We train the models described in Section~\ref{sec:models} using the datasets provided in Table~\ref{tbl:dataset_details}. To identify the best-performing model for each dataset, we conduct a comprehensive grid search and train the models for 50 epochs using three optimization algorithms: Stochastic Gradient Descent (SGD), Adam \cite{adaptive_momentum}, and AdamW \cite{adamw}. The initial learning rates are set to $10^{\{-5, -4, -3, -2\}}$. For SGD, we adopt a cosine annealing learning rate schedule, following the approach of~\cite{simsiam}. Additionally, we experiment with weight decays of $10^{-4}$, $10^{-5}$, and 0. All models are trained with a batch size of 32, and the models achieving the highest accuracy on the validation set are selected as the final models.

The experimental results, summarized in Table~\ref{tbl:results} provide a comprehensive evaluation of model performance across different dataset configurations. We report accuracy (Acc.), positive predictive value (PPV), and negative predictive value (NPV) for training, validation, and testing datasets. The baseline datasets without spurious correlations serve as a reference point, while the experimental datasets, which include spurious features, allow us to assess the impact of spurious correlations on model generalization.

\textbf{Consistent results with baseline datasets}.\,The baseline datasets show minimal performance degradation from validation to testing, with stable accuracy and predictive values across different dataset sizes. This consistency suggests that DNNs generalize well to unseen data when no spurious correlations are present, reinforcing the dataset’s validity.

\textbf{Weak or no spurious correlations}.\,While models trained on datasets containing spurious correlations on ethnicity, menopause status, and surgery type show some decline in testing accuracy, the impact is not severe. The test accuracy for these features remains above 70\%, indicating that although spurious correlations influence model predictions, their effect is not as dominant.

\textbf{Strong spurious correlations}.\,Models trained on datasets where magnetic field strength and vertical alignment are spuriously correlated with the target labels achieve near-perfect accuracy during training and validation but suffer a substantial drop in test performance, with accuracy declining to around 50\%. This sharp decrease indicates that models are heavily relying on these non-clinical attributes for decision-making rather than learning meaningful medical features. In the case of magnetic field strength, the model labels all 1.5T images as tumor-positive and all 3T images as tumor-negative, while for vertical alignment, images facing up are predicted as tumor-positive and those facing down as tumor-negative. This confirms that the model learns to exploit these non-clinical cues as shortcuts, failing to generalize when tested on unbiased data.

\textbf{Confirming strong spurious correlations}.\,
Based on the initial set of experiments provided above, we identify magnetic field strength and vertical alignment as the two features that, when spuriously correlated with image labels, lead models to learn these spurious signals instead of clinically meaningful features. To confirm these observations and ensure that the results presented in Table~\ref{tbl:results} are not merely one-off outcomes based on dataset sampling, we repeat the same experiment $10$ times with different randomized patients in the training, validation, and testing datasets. In all of those experiments, we find that validation accuracy reaches $\sim 100\%$ at first few epochs while test accuracy remains $\sim 50\%$. 

\begin{figure}[t!]
\centering
\begin{subfigure}{0.475\textwidth}
\includegraphics[width=0.24\textwidth]{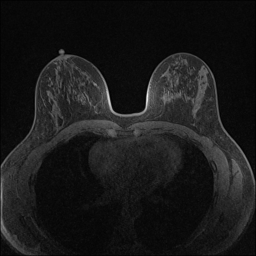}
\includegraphics[width=0.24\textwidth]{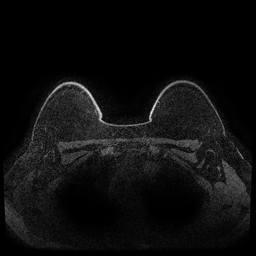}
\includegraphics[width=0.24\textwidth]{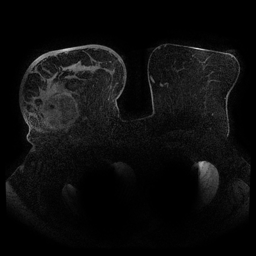}
\includegraphics[width=0.24\textwidth]{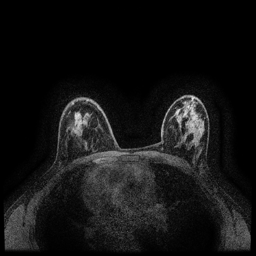}
\\
\includegraphics[width=0.24\textwidth]{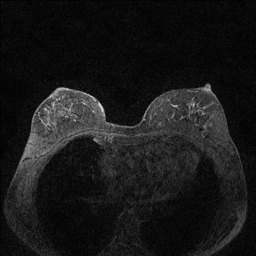}
\includegraphics[width=0.24\textwidth]{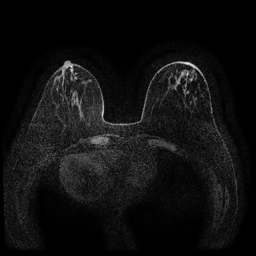}
\includegraphics[width=0.24\textwidth]{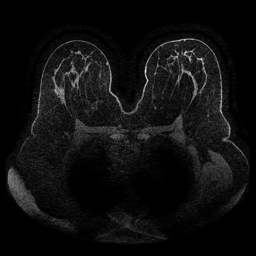}
\includegraphics[width=0.24\textwidth]{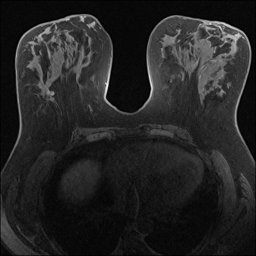}
\caption{MRIs from 1.5 Tesla device}
\end{subfigure}
\hspace{0.05em}
\begin{subfigure}{0.475\textwidth}
\includegraphics[width=0.24\textwidth]{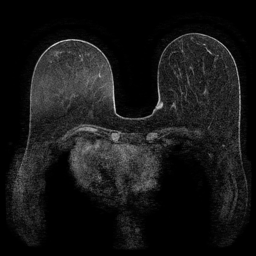}
\includegraphics[width=0.24\textwidth]{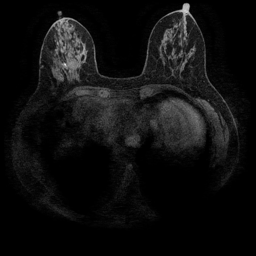}
\includegraphics[width=0.24\textwidth]{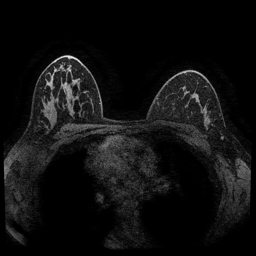}
\includegraphics[width=0.24\textwidth]{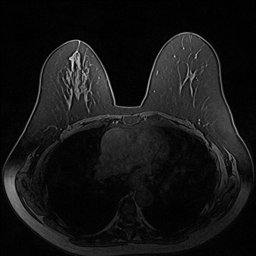}
\\
\includegraphics[width=0.24\textwidth]{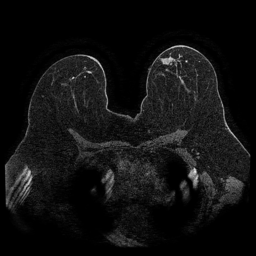}
\includegraphics[width=0.24\textwidth]{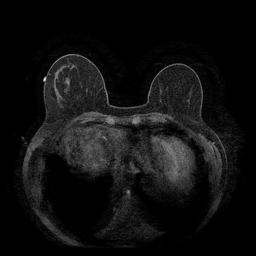}
\includegraphics[width=0.24\textwidth]{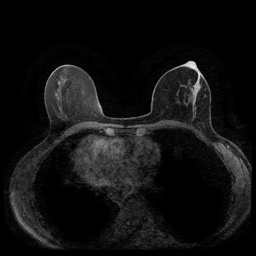}
\includegraphics[width=0.24\textwidth]{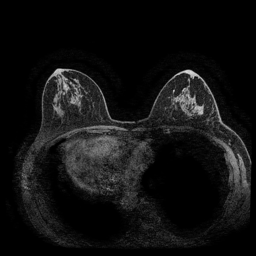}
\caption{MRIs from 3 Tesla device}
\end{subfigure}
\caption{Example breast MRI images obtained using (a) 1.5T and (b) 3T devices.}
\label{fig:mri_device_examples}
\end{figure}

\subsection{Understanding Spurious Correlations}

\hspace{1.75em}\textbf{Magnetic field strength}.\,3T scanners offer higher magnetic field strength, improving signal-to-noise ratio (SNR) and image resolution for sharper, more detailed images \cite{schmitt2004}. However, they are more prone to artifacts, heating effects, and signal loss, especially around metal implants \cite{graves2021}. As such, our proposed dataset involving this spurious signal features a non-local spurious signal that influences the entire image rather than a localized region. An example set of images obtained from 1.5T and 3T devices are provided in Figure~\ref{fig:mri_device_examples}, showing that it is visually not possible to distinguish 1.5T MRIs from the 3T ones.

\textbf{Vertical orientation}.\,Different from magnetic field strength, which affects the entire image globally, vertical orientation is a local feature that only alters the spatial arrangement of structures within the image. This transformation does not modify the underlying tissue characteristics or signal properties but instead introduces artificial correlations that models may exploit as shortcuts.

\section{Conclusions and Future Perspectives}

We introduce \textbf{SpurBreast}, a curated dataset designed to study the impact of spurious correlations in breast MRI classification. It includes two experimental datasets with specific biases to evaluate model robustness. The first dataset introduces a spurious correlation with MRI magnetic field strength and the second dataset introduces spurious correlations based on vertical alignment. In addition to these datasets with spurious correlations, we provide a baseline dataset free of spurious correlations, serving as a benchmark for unbiased evaluation and bias mitigation.

Our goal in providing datasets with spurious correlations is to enable researchers to investigate how models learn and rely on unintended features, measure uncertainty, and develop methods to improve model generalization. Models and datasets are available at \href{https://github.com/utkuozbulak/spurbreast}{github.com/utkuozbulak/spurbreast}.

\begin{credits}
\subsubsection{\ackname} The authors have no acknowledgments to declare.

\subsubsection{\discintname} The authors have no competing interests to declare that are relevant to the content of this article.
\end{credits}
\bibliographystyle{splncs04}
\bibliography{Paper-0408}

\end{document}